# Web Publishing of the Files Obtained by Flash

**Virgiliu Streian, Adela Ionescu**
**"Tibiscus" University, Timisoara, Romania**

**ABSTRACT.** The aim of this article is to familiarize the user with the Web publishing of the files obtained by Flash. The article contains an overview of Macromedia® Flash™ 5, as well as the running of a Playing Flash movie, information on Flash and Generator, the publishing of Flash movies, a HTLM publishing for Flash Player files and publishing by Generator templates.

## 1. General presentation of Macromedia® Flash™ 5

Macromedia® Flash™ 5, still refered to as Flash, is a product of the Macromedia Inc. Company, having its headquarters in San Francisco, California, USA.

On Internet, Macromedia is found at http://www.macromedia.com. Macromedia provides IT solutions to clients from the business, educational or governmental area. The company operates in more than 50 countries worldwide.

Flash is the software that enables millions of Internet application developers to find solutions targeted to Web users performed on different platforms and equipments.

## 2. Running Flash Playing Movies

Considering the theme of the present study we will emphasize the operation of publishing.

The flash publishing utility is used to display animation on the web. The publish commands create Flash Player files (SWF) and a HTML

document, which started from a browser, insert Flash Player files in a browser window.

The Flash Player format (SWF) is the main format for distributing a Flash content and it is the only format that supports all the interactive functionality of the Flash.

A Flash player movie can be played in several ways. The one taken into account by our study is made via some Internet browsers, such as Netscape Navigator or Internet Explorer, equipped with a Flash Player. The format of a Flash Player file is an open standard that is supported by other applications. See Macromedia site for latest information.

## 2.1 On Generator and Flash

Generator expands the utility area of Flash, enabling designers to create a Flash rich media content which can be finally delivered in a variety of graphic animation or static formats.

Any object created in Flash - including library elements, symbols, animations, published output - can be considered a Generator object, using the symbols or the Generator variables. (Generator Variables are text enclosed in curly brackets such as {text}.) Using Generator one can choose the best form to display the information - including scrolling lists, graphs, tables, a variety of graphic formats, sounds and movies - to create a customized multimedia Web experience in real time.

If Generator 2 is installed, then Flash can create templates that contain Generator variable (graphics, text and sound) which will replace the content from various data sources (text files, databases etc.). This generated content can be run in a client browser as a Flash Player movie or as an animated JPEG, PNG, GIF or QuickTime file.

## 2.2 Publishing Flash movies

Publishing a Flash movie on the web involves two phases. Firstly, all the files required by the application are prepared with the Publish Settings command. Then publish the movie and all the corresponding files are published with the *publish command*. In order to prepare a Flash content to be used in other applications one can use an *export* command.

The Publish Settings command allows the user to choose the formats and to specify the settings for individual files containing movies - including GIF, JPEG or PNG - and to store these settings with the movie file.

Depending on what it was specified in the Publish Settings dialog box, a *publish command* will create the following files:

- Flash movie for Web (SWF file).
- Alternative images in a variety of formats that automatically appear if Flash Player is not installed (GIF, JPEG, PNG and QuickTime).
- The support of the HTML document necessary to display a movie (or alternative image) in the browser and the control of the browser settings.
- Stand-alone Projectors both for Windows and Macintosh systems and QuickTime videos for Flash movies (EXE, HQX, respectively MOV files).

In order to change a Flash Player movie created with the original Flash movie should be edited. Afterwards a new *publish command* is made. Importing a Flash Movie Player in Flash determines the loss of some information.

## 2.3 Publishing HTML file Flash Player

Running a Flash movie into a Web browser requires a HTML document to run the movie under some specified browser settings conditions. This document is automatically generated by a *publish* command, starting from the HTML parameters stored in a template document. HTML parameters determine where a Flash movie appears in the browser window, the background color (background color), the film size, etc.., and it sets the attributes for the OBJECT and EMBED tags. All these parameters can be change in the HTML panel from the Publish Settings dialog box. Changing these settings covers the options that were originally set in Flash movie. The changed settings are inserted in the template document. The template document can be any text file containing the correspondent template variables - including a simple HTML file, which may contain a code specially interpreted, for example, the ColdFusion or Active Server Pages (ASP), or a template included by Flash (for more information see *Presentation of HTML templates for publication*).

A template can be adapted (see *Customizing a HTML publishing template*) or manually modified (input of HTML parameters necessary to run Flash movies) with any HTML editor (see *Editing Flash HTML Settings).*

## 3. Publishing via Generator templates and the description of HTML publishing templates

Generator provides the possibility of adding a dynamic content such as text, graphics and sound to a Flash movie. The publishing options can be specified in the Generator panel from the Publish Settings box.

The Flash HTML templates enable a checking such as what type of the film is taken into account on the Web and how it is like and how it runs in the Web browser. A Flash template is a text file containing both a fix HTML code and template code (variable). When a Flash movie is published, Flash replaces the variables from the template with the selected values in the Publish Settings box - HTML settings, and produces an HTML page with embedded video.

Flash includes several templates covering the needs of many users, which virtually eliminates the need of editing a HTML page with Flash film. For example, a simple template places a Flash movie into a HTML page generated so that the user could see through a Web browser if the plug-in is installed. Other templates firstly detect whether the plug-in has been installed and if not, they install it.

One can easily use an existing template, change settings, and publish a new HTML page. HTML experts can create their own templates using any HTML editor. Creating a template is identical to creating a standard HTML page, except for the fact that the variables (starting with $ sign) specific to Flash movies can be replaced with values.

Flash HTML templates have the following characteristics:
- A single headline that appears on the Template pop-up menu
- A longer description that appears when you click the Info button
- Template variables starting with $ sign indicating that they will be replaced when Flash generates output file (\ $ is used if the use of $ sign is necessary for other purposes).
- HTML tags, OBJECT and EMBED, comply with requirements of Microsoft Internet Explorer, respectively Netscape Communicator / Navigator. In order to properly display a movie in a HTML page the tag requirements must be respected. Internet Explorer opens a Flash movie using HTML tag OBJECT; while Netscape HTML uses the tag EMBED. For more information see *Use of OBJECT and EMBED*.

### 3.1 Customizing a publish HTML template

The users of HTML language can change the HTML template variables to create map images, or text reports or URL reports to insert their own values for the common Flash parameters OBJECT and EMBED (for Internet Explorer, respectively Netscape Communicator / Navigator).

Flash templates can include any HTML content necessary to an application or even a code for special interpreters such as Cold Fusion, ASP and other such.

### 3.2 Creating an image map (sensitive map)

Flash can generate an image map using any image. Flash inserts the code for the image map as for a virtual image, so that, if the reference image is replaced, the map feature remains. The $IU variable identifies the name of the GIF files, JPEG or PNG file.

To create an image map:

1. For a Flash movie, the keyframe used to label the map image # map in the Frame panel (Windows> Panels> Frame). You can use any keyframe with buttons which have attached Get URL actions.

   If no frame label is created, Flash creates an image map using the buttons from the last frame or film. This option generates an embedded

2. Map image, not an embedded Flash movie.

3. From an HTML editor, a HTML template to be modified is opened. Flash stores the HTML templates in the Macromedia Flash 5/HTML directory.

4. Save the template.

5. Select File> Publish Settings, click the Format tab and select a format for an image map-GIF, JPEG or PNG.

6. Click OK to save settings.

For example the following code can be inserted in a template:

```
$IM

```

That will create the following code in a HTML document created through a Publish command:

```
<MAP NAME="mymovie">
```

```
<AREA COORDS="130,116,214,182"
HREF="http://www.macromedia.com">
</MAP>
<IMG SRC="mymovie.gif" usemap="#mymovie" WIDTH=550
HEIGHT=400 BORDER=0>
```

## 3.3 Creating a text report

The $MT variable template determines Flash to insert all the texts in the current Flash movie as a HTML comment code. This is very useful for indexing the content of a film and for making it visible to search engines.

## 3.4 Creating a URL

The $MU variable template determines Flash to generate a list of URLs that relates actions in the current movie and inserts it in a current location as a comment. This validates a mechanism for checking the links to see and check the links from a film.

## 3.5 The use of template variables

The $PO template variables (for the OBJECT tags)) or $ PE (for EMBED tags) are some of the most commonly used. Both variables make Flash to insert into a template non implicit values for a series of parameters of the OBJECT or EMBED tags, such as PLAY ($PL), QUALITY ($QU), SCALE ($SC), SALIGN ($SA), WMODE ($WM), DEVICEFONT ($DE) or BGCOLOR ($BG) or in the case of OBJECT they appear in such lines: OBJECTIVE tags or embed such as PLAY ($ PL), QUALITY ($ qu), SCALE ($ SC) SALIGN ($ SA) WMODE ($ WM) DEVICEFONT ($ DE) or BGCOLOR ($ BG) which in his case OBJECTIVE appear in such lines:

```
<PARAM NAME="PLAY" VALUE="true">
<PARAM NAME="QUALITY" VALUE="high">
...etc...
```

See *Sample of template* section below for examples of these variables.

### 3.6 Example of template

The default template Default.html from Flash is taken as an example that
contains the use of several variables template.

```
$TTFlash Only (Default)
$DS
Use an OBJECT and EMBED
tag to display Flash.
$DF
<HTML>
<HEAD>
<TITLE>$TI</TITLE>
</HEAD>
<BODY bgcolor="$BG">
<!-- URLs used in the movie-->
$MU
<!-- text used in the movie-->
$MT
<OBJECT classid="clsid:D27CDB6E-AE6D-11cf-96B8-
444553540000"
codebase="http://download.macromedia.com/pub/shockwave/
cabs/flash/swflash.cab#version=5,0,0,0"
 ID=$TI WIDTH=$WI HEIGHT=$HE>
 $PO
<EMBED $PE WIDTH=$WI HEIGHT=$HE
 TYPE="application/x-shockwave-flash"
PLUGINSPAGE="http://www.macromedia.com/shockwave/downlo
ad/index.cgi?P1_Prod_Version=ShockwaveFlash"></EMBED>
</OBJECT>
</BODY>
</HTML>
```

## 4. Editing Flash HTML settings

In order to run a Flash movie into a web browser a HTML document that
specifies the browser settings is necessary. For those with experience in HTML,
the parameters may be changed, or manually introduced using a HTML editor or
their own HTML files can be created to control a Flash movie.

For information about how Flash automatically creates an HTML
document when publishing a movie see *Publish a movie in Flash*. For more
information about customizing HTML templates included in Flash see
*Customizing a template publication HTML.*

**4.1 Using OBJECT and EMBED**

For a Flash Player to display a movie in a Web browser, a HTML document must use OBJECT and EMBED tags with the corresponding parameters.

For OBJECT there are four settings (HEIGHT, WIDTH, CLASSID and CODEBASE) appearing in the OBJECT tag; the rest represents parameters that appear in separate PARAM. tags. For example:

```
<OBJECT          CLASSID="clsid:D27CDB6E-AE6D-11cf-96B8-
444553540000"                    WIDTH="100"HEIGHT="100"
CODEBASE="http://active.macromedia.com/flash5/cabs/swfl
ash.cab#version=5,0,0,0">
<PARAM NAME="MOVIE" VALUE="moviename.swf">
<PARAM NAME="PLAY" VALUE="true">
<PARAM NAME="LOOP" VALUE="true">
<PARAM NAME="QUALITY" VALUE="high">
</OBJECT>
```

For EMBED tag, all the settings (such as HEIGHT, WIDTH, QUALITY and LOOP) are attributes which appear between angular brackets of the EMBED tag. For example:

```
<EMBED  SRC="moviename.swf"  WIDTH="100"  HEIGHT="100"
PLAY="true"          LOOP="true"          QUALITY="high"
PLUGINSPAGE="http://www.macromedia.com/shockwave/downlo
ad/index.cgi?P1_Prod_Version=ShockwaveFlash">
</EMBED>
```

In order to use both tags, the tag EMBED is posed exactly before closing the OBJECT tag, such as:

```
<OBJECT        CLASSID="clsid:    D27CDB6E-AE6D-11cf-96B8-
444553540000"            WIDTH="100"            HEIGHT="100"
CODEBASE="http://active.macromedia.com/flash5/cabs/swfl
ash.cab#version=5,0,0,0">
<PARAM NAME="MOVIE" VALUE="moviename.swf">
<PARAM NAME="PLAY" VALUE="true">
<PARAM NAME="LOOP" VALUE="true">
<PARAM NAME="QUALITY" VALUE="high">
<EMBED  SRC="moviename.swf"  WIDTH="100"  HEIGHT="100"
PLAY="true"          LOOP="true"          QUALITY="high"
PLUGINSPAGE="http://www.macromedia.com/shockwave/downlo
ad/index.cgi?P1_Prod_Version=ShockwaveFlash">
</EMBED>
</OBJECT>
```

Note: If both OBJECT and EMBED tags are used then identical values must be used for each attribute or parameter in order to provide the consistence of the movie running no matter the used browser.

The swflash.cab # version = 5,0,0,0 parameter is optional and it can be omitted if the control of the versions number is not wanted.

The following tag attributes and parameters appear in HTML tag created by a PUBLISH command. The list below may be referred to when the manual writing of an own HTML inserting a Flash movie is wanted. Except for those mentioned above, all items below apply to both OBJECT and EMBED tags. The optional inputs are specified. When a template is customized a variable template can be replaced.

The list of attributes and parameters that can occur in an EMBED or OBJECT tag:

| Attribute | Value | Variable template |
|---|---|---|
| *SRC* | `movieName.swf` | `$MO` |
| *MOVIE* | `movieName.swf` | `$MO` |
| *CLASSID* | `clsid:D27CDB6E-AE6D-11cf-96B8-444553540000` | |
| *WIDTH* | `n or n%` | `$WI` |
| *HEIGHT* | `n or n%` | `$HE` |
| *CODEBASE* | `http://active.macromedia.com/flash5/cabs/swflash.cab#version=5,0,0,0"` | |
| *PLUGINSPAGE* | `http://www.macromedia.com/shockwave/download/index.cgi?P1_Prod_Version=ShockwaveFlash` | |
| *SWLIVECONNECT* | `true │ false` | |
| *PLAY* | `true │ false` | `$PL` |
| | | `$LO` |
| *QUALITY* | `low │ high │ autolow │ autohigh │ best` | `$QU` |
| *BGCOLOR* | `#RRGGBB` (hexadecimal RGB value) | `$BG` |
| *SCALE* | `showall │ noborder │ exactfit` | `$SC` |
| *ALIGN* | `L │ R │ T │ B` | `$HA` |
| *SALIGN* | `L │ R │ T │ B │ TL │ TR │ BL │ BR` | `$SA` |
| *BASE* | `base directory or URL` | |
| *MENU* | `true │ false` | `$ME` |
| *WMODE* | `Window │ Opaque │ Transparent` | `$WM` |

**5. A Web server Flash configuration for Flash**

To access a file from a Web server, in order to be displayed, the server must properly identify that it is a Flash Player file. If on the MIME type data server it misses or it is not configured correctly, your browser may display an error message or a window containing a lot of icons.

To test the server's configuration, see TechNote # 12696 on Macromedia Flash Support Center, http://www.macromedia.com. If the server is not correctly configured the MIME Flash Player must be added in the configuration files of the server and the following Flash Player file extensions must be associated.

MIME type application / x-shockwave-flash have the file extension. Swf.

MIME type application / futuresplash have the extension file.spl.

If the server is a Macintosh, you should also set the following parameters: Action: Binary; Type: SWFL and Creator: SWF2.